\documentclass[prb,twocolumn]{revtex4}
\usepackage{graphicx}
\usepackage{amsmath}
\newcommand{\smfrac}[2]{\mbox{$#1 \over #2$}}
\newcommand{\bxo}{{\bf x}_0}
\newcommand{\bxodot}{\dot{\bf x}_0}

\newcommand{\bx}{{\bf x}}
\newcommand{\bA}{{\bf A}}
\newcommand{\bAf}{{\bf A}_{\rm ext}}

\newcommand{\bB}{{\bf B}}
\newcommand{\bBb}{{\bf B}_b}
\newcommand{\bBf}{{\bf B}_{\rm ext}}
\newcommand{\bD}{{\bf D}}
\newcommand{\bE}{{\bf E}}
\newcommand{\bEb}{{\bf E}_b}
\newcommand{\bEf}{{\bf E}_{\rm ext}}

\newcommand{\Fd}{F_{\rm diel}}
\newcommand{\FLand}{F_{\rm L}}
\newcommand{\Fm}{F_{\rm mat}}
\newcommand{\Fmprime}{{F^\prime_{\rm mat}}}
\newcommand{\Fcomb}{F_{\rm comb}}
\newcommand{\Ham}{{\cal H}}
\newcommand{\Hcomb}{{\cal H}_{\rm comb}}
\newcommand{\Hamd}{{\cal H}_{\rm diel}}
\newcommand{\Hamm}{{\cal H}_{\rm mat}}
\newcommand{\Hammprime}{{\cal H'}_{\rm mat}}
\newcommand{\bH}{{\bf H}}
\newcommand{\bJ}{{\bf J}}
\newcommand{\bJb}{{\bf J}_b}
\newcommand{\bJf}{{\bf J}_{\rm ext}}
\newcommand{\bM}{{\bf M}}
\newcommand{\bm}{{\bf m}}
\newcommand{\bnabla}{\boldsymbol{\nabla}}
\newcommand{\bPi}{\boldsymbol{\Pi}_{\bf A}}
\newcommand{\bP}{{\bf P}}
\newcommand{\phib}{{\phi_b}}
\newcommand{\phif}{\phi_{\rm ext}}
\newcommand{\bp}{{\bf p}}
\newcommand{\bpo}{{\bf p}_0}

\newcommand{\rhob}{\rho_{b}}
\newcommand{\bs}{{\bf s}}
\newcommand{\rhof}{\rho_{\rm ext}}
\newcommand{\Tr}{\rm Tr\,}
\begin{document}
\title{Free energies in the presence of electric and magnetic fields}
\author{Onuttom Narayan}
\email{narayan@physics.ucsc.edu}
\author{ A.~P.~Young}
\affiliation{Department of Physics,
University of California,
Santa Cruz, California 95064}

\begin{abstract}
We discuss different free energies for materials in static
electric and magnetic
fields.  We explain what the corresponding Hamiltonians are, 
and describe which choice gives rise to which result for the free
energy change, $d F$, in the thermodynamic identity. We also discuss which
Hamiltonian is the most appropriate for calculations using
statistical mechanics, as well as the relationship between the
various free energies
and the ``Landau function'', which has to be minimized to determine
the equilibrium polarization or magnetization, and is central to Landau's
theory of second order phase transitions.

\end{abstract}

\maketitle
\section{Introduction}

One frequently needs to calculate the properties of materials in magnetic or
electric fields using statistical mechanics. The procedure, of course,
is to start from ``the Hamiltonian'', perform the statistical sum to get the
partition function, the logarithm of which gives ``the free energy''.
Unfortunately, there is much confusion in the literature because different
authors include different pieces of the energy of the electric or magnetic
field in the Hamiltonian, generally without explaining precisely what they
are doing. In this paper we describe the different results that are given
for the cases of
static electric fields and static magnetic fields, show the relations
between them, and explain which choice gives rise to which result for the free
energy change $d F$ in the thermodynamic identity. We also discuss which
Hamiltonian, and hence free energy, is the most appropriate. Finally we
describe the connection between the various free energies and the ``Landau
function'' which has to be minimized in Landau's theory of second order phase
transitions. 

We use rationalized Gaussian units throughout, since this encumbers the
formulae with a minimum number of factors for units and numerical factors. We
also set $\hbar = c = 1$.

\section{Lagrangian and Hamiltonian}
For a charged particle in an electromagnetic field
the Lagrangian is 
\begin{equation}
L = {1\over 2} m \bxodot^2 + q \bxodot \cdot \bA(\bxo)
- q \phi(\bxo) + {1\over 2}\int d{\bx}
\left[\bE^2 - \bB^2\right] \, ,
\label{Lagrangian}
\end{equation}
where $\bxo$ and $\bxodot$ are the position and velocity of the particle
which has charge $q$, and the scalar potential $\phi$ and vector potential 
$\bA$ are related to the electric and magnetic fields by
$\bE = -\bnabla \phi - \partial_t\bA$ and
$\bB  = \bnabla \times \bA.$ This parametrization automatically ensures
that $\bnabla\times\bE = 0$ and $\bnabla\cdot\bB = 0.$
The justification for this Lagrangian is that the Euler-Lagrange
equations of motion derived
from it for $\phi,$ $\bA$ and $\bxo$ correctly give the remaining
two (inhomogeneous) Maxwell's equations, with charge density and current
density given by
\begin{equation}
\rho(\bx) = q\, \delta(\bx - \bxo)  \, , \qquad \bJ(\bx)
= q\, \bxodot\, \delta(\bx
- \bxo)  \, ,
\label{rho}
\end{equation}
as well as the equation of motion for the particle under the Lorentz
force~\cite{eofm}.

For statistical mechanics we need the Hamiltonian, and
to obtain to this from the Lagrangian in Eq.~(\ref{Lagrangian}), we 
first have to construct the canonical momenta. 
There is no canonical momentum for $\phi$ since $L$ does not depend on
$\partial_t \phi$; $\phi$ is not a dynamical variable. For the other degrees 
of freedom, we have 
$\bpo\equiv \partial L / \partial \bxodot 
= m \bxodot + q \bA$ and $\bPi
\equiv \partial L / \partial (\partial_t \bA)
= \bnabla \phi + \partial_t \bA \ = -\bE.$
Hence the Hamiltonian is equal to 
\begin{eqnarray}
\Ham & = & \bpo \cdot \bxodot + \int d\bx
\left( \bPi \cdot \partial_t \bA\right) - L \nonumber \\
& = & 
{1\over{2m}}\left(\bpo - q \bA(\bxo)\right)^2 + q \phi(\bxo) \nonumber \\
&   & \ +\int d\bx \left\{ {1\over2} \left[ \left(\bnabla \times \bA\right)^2  +
\bPi^2 \right]  - \bnabla \phi \cdot \bPi \right\} \, .
\label{Hamiltonian}
\end{eqnarray}
Applying
Hamilton's equations of motion to Eq.~(\ref{Hamiltonian}) gives Maxwell's
equations and the equation of motion for the particle under the Lorentz force,
as required.
Equation~(\ref{Hamiltonian})
can be simplified by noting that since $\phi$ is not a dynamical variable,
we can insert the condition 
\begin{equation}
\bnabla\cdot\bE = \rho
\label{divE}
\end{equation}
corresponding to 
$\partial_\phi L = - \partial_\phi\Ham = 0,$ in the Hamiltonian itself. 
Since $\bPi = -\bE$,
the last term in Eq.~(\ref{Hamiltonian}) can
be rearranged as
\begin{equation}
\int d\bx\, \bnabla \phi \cdot \bE  =   -\int d\bx\, \phi \bnabla \cdot \bE
 =  -\int d\bx\, \phi \rho \nonumber 
= - q \phi(\bxo) \, ,
\end{equation}
in which we integrated by parts in the first step, used Eq.~(\ref{divE})
in the second step, and used Eq.~(\ref{rho}) in the last step. Hence this term
cancels the $+q \phi(\bxo)$ term in the Hamiltonian,
Eq.~(\ref{Hamiltonian}). The Hamiltonian can therefore
be simplified to
\begin{equation}
\Ham = 
{1\over{2m}}\left(\bpo - q \bA(\bxo)\right)^2  - \gamma\bs \cdot \bB +
{1\over 2} 
\int d\bx \, \left[ \bE^2 + \bB^2 
\label{simpleham}
\right]   \, ,
\end{equation}
in which the condition $\bnabla \cdot \bE = \rho$ must be imposed, and we have
included the magnetic coupling to the particle's spin $\bs$ with $\gamma$ being 
the gyromagnetic ratio.

In the next two sections we consider the problem of a many-particle
\textit{system} of charged particles in the presence of an external field
produced by some \textit{external apparatus}. An
important task will be to separate out
which pieces of the Hamiltonian should be associated with the system and which
with the apparatus. 

\section{Electrostatics}
In this section we neglect the effects of the
magnetic field and set
\begin{equation}
\bA = 0\quad  (\mbox{so }\bB = 0) \, , \ \quad
\bnabla \times \bE = 0\, ,\  \quad \bE = - \bnabla \phi \, ,
\end{equation}
where $\phi$ is independent of time.  We
consider a situation in which there is a dielectric which
has bound charge density $\rhob$ giving rise to an electric
field $\bEb$, as well as free charges $\rhof$, entirely outside the dielectric, which
give rise to an external field $\bEf$. We have
\begin{equation}
\rho = \rhob + \rhof \, , \qquad
\bE = \bEb + \bEf \, ,
\label{EfplusEb}
\end{equation}
and the relation between charge density and corresponding electric field is
given by Gauss' law
\begin{equation}
\bnabla \cdot \bE = \rho\, , \qquad 
\bnabla \cdot \bEb = \rhob\, , \qquad \bnabla \cdot \bEf = \rhof \, .
\label{Ebf}
\end{equation}

To obtain the Hamiltonian of the dielectric,
we remove
$(1/2)\int d\bx  \, \bEf^2$ from Eq.~(\ref{simpleham}), since this is
the work done to set up the free charges, and should be considered part of the
energy of the
external apparatus, not the dielectric.
Hence the Hamiltonian of the dielectric is given by
\begin{eqnarray}
\Hamd & = &  \Ham_0 + 
{1 \over 2} \int d\bx \, \left[ \bE^2 - \bEf^2 \right]
\label{HamdE}\\
& = &
\Ham_0 + \Ham_1 + \Ham_2  \, ,
\label{Hamd}
\end{eqnarray}
where $\Ham_0$ is the part of the Hamiltonian not \textit{explicitly}
dependent on the electric field, $\Ham_1$ is given by
\begin{equation}
\Ham_1  =  
\int d\bx \, \bEb \cdot \bEf \, ,
\label{Ham1}
\end{equation}
and
\begin{equation}
\Ham_2  =  
{1\over 2} \int d\bx \, \bEb^2 \, .
\label{Ham2}
\end{equation}
Note that there is an \textit{implicit} dependence of $\Ham_0$ on the electric
field because the interactions in $\Ham_0$
depend on the locations of the particles, and these affect the
field through Gauss' law Eq.~(\ref{Ebf}). 

It is useful to express Eqs.~(\ref{Ham1}) and (\ref{Ham2}) in a different way.
Writing $\bEf = - \bnabla \phif$ and integrating by parts,
Eq.~(\ref{Ham1}) can be written 
\begin{equation}
\Ham_1 = \int_V d\bx\, \rhob \phif \, ,
\label{Ham1b}
\end{equation}
showing that $\Ham_1$ represents the interaction between charges in the
dielectric $\rhob$
and the external electric potential $\phif$. The integral in Eq.~(\ref{Ham1b})
is only over the volume $V$ of the sample, and in the rest of the paper 
we will find it convenient to
indicate integrals restricted to the volume of the sample by a subscript $V$.
Integrals with no such indicator will be
over all space.
Similarly Eq.~(\ref{Ham2}) can be written
\begin{equation}
\Ham_2 = {1 \over 2} \int_V d\bx\, \rhob \phib = 
{1\over 2} \int_V d\bx_1 \int_V d\bx_2 \, {\rhob(\bx_1) \rhob(\bx_2) \over
4 \pi |\bx_1 - \bx_2|} 
\, ,
\label{Ham2b}
\end{equation}
showing that $\Ham_2$ represents the self-interaction of charges in the 
dielectric. In the second expression in Eq.~(\ref{Ham2b}) we used Coulomb's
law.

It is conventional to describe the response of a dielectric to an external
field by its polarization $\bP$, satisfying 
$\bnabla \cdot \bP = -\rhob.$ 
The ``electric displacement field" is defined
by $\bD = \bE + \bP,$ so that $\bnabla \cdot \bD = \rhof.$ Now
any vector field
(with appropriate boundary conditions at infinity) can be
written~\cite{griffiths,jackson}
as the sum
of a longitudinal part (which has zero curl) and a transverse part (which has
zero divergence). Comparing the equations for $\bnabla\cdot\bP$ and 
$\bnabla\cdot\bD$ with with Eq.~(\ref{Ebf}), we have
\begin{equation}
\bP^L = -\bEb \, ,\qquad\qquad \bD^L = \bEf \, ,
\label{PL}
\end{equation}
where $\bP^L$ and $
\bD^L$ are the longitudinal parts of $\bP$ and $\bD$ respectively.
However, $\bP \ne -\bEb$ (and $\bD \ne \bEf$) because $\bP$ also has a 
transverse component $\bP^T$, whereas the electric field is purely
longitudinal since $\bnabla \times \bE = 0$ in an electrostatic situation.
Note that even with the physical condition that $\bP$ must vanish outside
the material, $\bP^T$ is ambiguous, although it does not result in any (bound) 
charges and hence has no measurable consequences. If the polarization
is a local effect, with dipole moments being induced in response to the local
electric field, one can impose the additional condition ${\bf P} = f({\bf E})$
(in the simplest case, a linear function) to determine the polarization
uniquely. However, when polarization is a non-local process, e.g. in a
conductor, it is genuinely ambiguous. For example, for an infinitely long
cylindrical conductor of radius $a$, one can add an azimuthal polarization
depending only on $r$ [i.e.
$P_\theta = {\rm const.}\, g(r) \, $ for $0 < r < a$ and
$\bP = 0 $ for $ r > a$],
without changing the electric field.

Now
the integral over all space of the product of a longitudinal field and a
transverse field is zero, which is easily proved by writing the
longitudinal field as the gradient of a scalar potential and integrating
by parts.
Hence, since $\bP=0$ outside the sample,
\begin{eqnarray}
\int_V d\bx \, \bP \cdot \bEf & = & 
\int d\bx \, \bP \cdot \bEf = \int d\bx \, \bP^L \cdot \bEf \nonumber \\
& = &  
-\int d\bx \, \bEb \cdot \bEf \, .
\end{eqnarray}
Consequently $\Ham_1$ in Eq,~(\ref{Ham1}) can be reexpressed as
\begin{equation}
\Ham_1  =  
- \int_V d\bx \, \bP \cdot \bEf \, .
\label{Ham1c}
\end{equation}

Hence, putting all this together the Hamiltonian of the dielectric is given by
\begin{eqnarray}
\Hamd & = & 
\Ham_0
- \int_V d\bx\, \bP \cdot \bEf  \nonumber \\
& & \ +{1 \over 2} \int_V d\bx_1 \int_V d\bx_2\,
{\rhob(\bx_1) \rhob(\bx_2) \over
4 \pi |\bx_1 - \bx_2|}  \, .
\label{Hamd_all}
\end{eqnarray}
The free energy is then given by
\begin{equation}
\Fd = -k_B T \ln \Tr e^{-\beta\Hamd} \, .
\end{equation}
where the trace is over the states of the system.

In equilibrium, 
the free energy is minimized with respect to $\rhob$. Let us now change the
external field by some small amount $\delta \bEf$. The bound charges will
respond but this causes no change in $\Fd$ to first order because $\Fd$ is at
a minimum with respect to $\rhob$. Hence the change in $\Fd$
comes entirely from the dependence
of $\Ham_1$ on $\bEf$, (see Eq.~(\ref{Ham1c})) which gives\cite{Kittel}
\begin{equation}
\delta \Fd = -\int_V d\bx\,  \bP \cdot \delta \bEf \, ,
\label{dFdE}
\end{equation}
which is a \textit{thermodynamic}
definition of the polarization. The integral is, of course, only over the
sample, not all space, because $\bP$ is only non-zero in the sample.
If instead one moves the dielectric with $\bEf$ fixed, Eq.~(\ref{dFdE})
computed in the reference frame of the dielectric also gives the work done in
moving the dielectric\cite{Kittel}. 
In situations where $\bP(\bx)$ can not be defined unambiguously, it may be
more convenient to replace $\bP$ by $-\bEb$ and integrate over all space.

Frequently, e.g. Pippard~\cite{pippard},
a different result for the change in free
energy is given, but this corresponds to a 
different free energy which 
omits the piece $\Ham_1$ of the Hamiltonian, 
the interaction between the dipole moment of the
system and the external field. Denoting the modified Hamiltonian by
$\Hamd'$ we have
\begin{equation}
\Hamd' = \Ham_0 + {1 \over 2} \int d\bx\, \bEb^2\, ,
\end{equation}
which consists of all terms that do not explicitly involve the
\textit{external}
electric field. Now $\Hamd - \Hamd' =
- \int_V d\bx\, \bP \cdot \bEf$, and the relationship between $\Fd$ and
$\Fd'$ is the same\cite{Kittel},
\begin{equation}
\Fd = \Fd' - \int_V d\bx\, \bP \cdot \bEf  \, ,
\label{legendre_el}
\end{equation}
i.e. a standard Legendre transformation, and so,
using Eq.~(\ref{dFdE}), we have\cite{pippard,Kittel}
\begin{equation}
\delta \Fd'  =  
\int_V d\bx\, 
\bEf \cdot \delta \bP \, .
\label{dFdP}
\end{equation} 
Thus $\Fd'$ (which depends upon the position of the charges) should be
regarded as a function of the polarization $\bP$, and its derivative with
respect to $\bP$ gives the external electric field.

It is also interesting to consider the free energy corresponding to including
just the Hamiltonian $\Ham_0$. Similar arguments to those given above yield
\begin{equation}
\delta F_0 = \int_V d\bx\,
\bE \cdot \delta \bP \, .
\label{dF0dP}
\end{equation}

The last free energy that we consider in this section is the
\textit{combined} free energy
of the external apparatus and dielectric. From Eq.~(\ref{simpleham})
and Eq.~(\ref{HamdE}), 
this is given by
\begin{equation}
\Fcomb
= F_0 + {1\over 2} \int d\bx\, \bE^2  = \Fd + {1\over 2} \int d\bx\, \bEf^2\, .
\label{Fcomb}
\end{equation}
Using Eq.~(\ref{dFdE}), a small change in the fields gives
\begin{eqnarray}
\delta \Fcomb   & = & \int d\bx\,(\bEf - \bP)\cdot\delta\bEf \nonumber \\
& = & \int d\bx\,(\bD - \bP)\cdot\delta\bEf =
\int d\bx\, \bE\cdot\delta\bEf\nonumber\\
& = & 
\int d\bx\, \bE \cdot \delta \bD  \, .
\label{dFsadD}
\end{eqnarray}
At the first step, the integral of $\bP\cdot\delta\bEf$ is extended to cover
\textit{all} space. At the second and last steps, we use Eq.~(\ref{PL}) and 
the fact that $\delta\bEf$ and $\bE$ are longitudinal. 
Equation (\ref{dFsadD})
is the total
work done on the \textit{free} charges in setting up the field in the
presence of the dielectric and is the form usually presented in the E\&M
books such as
Jackson~\cite{jackson}
or Griffiths~\cite{griffiths}.

\begin{table*}
\begin{center}
\begin{tabular}{|l|l|}
\hline\hline
\hspace{2cm}Hamiltonian &
\hspace{0.8cm}Free energy change \\
\hline\hline
$ \displaystyle \Hamd 
= \Ham_0 +\smfrac{1}{2}\int d\bx\,(\bEb^2 + 2 \bEb\cdot \bEf) $ &
$\displaystyle \delta \Fd \  \,= - \int_V d\bx\, \bP \cdot
\delta \bEf$~[\onlinecite{Kittel}]\\
\hline
$ \displaystyle \Hamd' 
= \Ham_0 +\smfrac{1}{2}\int d\bx\,\bEb^2 $ &
$\displaystyle \delta \Fd' \  \,=  \int_V d\bx\, \bEf \cdot
\delta \bP $~[\onlinecite{Kittel,pippard}] \\
\hline
$\displaystyle \Hcomb
= \Ham_0 +\smfrac{1}{2}\int d\bx\,(\bEb^2 + 2 \bEb\cdot \bEf + \bEf^2) $  &
$\displaystyle \delta \Fcomb = 
\int d\bx\, \bE \cdot \delta \bD
$~[\onlinecite{jackson,griffiths}] \\
\hline
$\displaystyle \Ham_0 $ &
$ \displaystyle \delta F_0 \quad\ = \int_V d\bx\,
\bE \cdot \delta \bP$\\
\hline\hline
\end{tabular}
\caption{Different free energies for the electrostatic case. For each entry,
the
free energy is related to the given Hamiltonian by $F = -k_B T \ln \Tr
e^{-\beta\Ham}$. We have written the electric field piece of the energy in
terms of the separate contributions from the bound and free charges.
In this way we clearly see that
the different Hamiltonians and free energies just
differ as to which pieces of the electric field energy,
$(1/2)\int d\bx\, (\bEb^2 + 2 \bEb \cdot \bEf + \bEf^2)$, are included.
The most useful form is the first line, since it includes all
pieces of the Hamiltonian involving the system. Furthermore, the expression
for $\delta \Fd$ also gives the
work done on the 
dielectric if it is moved in the external field. 
\label{table_el}
}
\end{center}
\end{table*}
Eqs.~(\ref{dFdE}), (\ref{dFdP}), (\ref{dF0dP}) 
and (\ref{dFsadD}) describe
different thermodynamic
derivatives in the presence of an electric field, each of which
involves a different
free energy depending upon which pieces of the electric field energy
$(1/2)\int d\bx\, (\bEb^2 + 2 \bEb \cdot \bEf + \bEf^2)$ are included.
These are summarized in Table~\ref{table_el}. When the definition of 
$\bP,$ $\delta\bP$ and $\bD$ are ambiguous, one can replace them with 
$\bEb,$ $\delta\bEb$ and $\bEf$ 
respectively in the final column of Table~\ref{table_el}, with the integrals
over all space. This is because of Eq.~(\ref{PL}) and the fact that 
$\delta\bEf,$  $\bEf$ and $\bE$ are longitudinal.
Of the free energies in Table~\ref{table_el},
$\Fd$ is the most useful since it includes everything
associated with the dielectric: (i) the
internal interactions not involving 
the electric field, (ii) the coupling of the charges in the
dielectric to an external field, and (iii) the Coulomb interaction between
charges in the dielectric, see Eq.~(\ref{Hamd_all}).
Hence it is surprising that Eq.~(\ref{dFdE}) seems to be
less often quoted than Eq.~(\ref{dFdP}).

\section{Magnetostatics}
In this section we consider a static magnetic field and so set 
\begin{equation}
\bE = 0 \, , \qquad \bB = \bnabla \times \bA \, .
\end{equation}
The
``system'' is a
magnetic material with ``bound'' current density 
$\bJb$ producing a magnetic field $\bBb$, and in addition there is
a source of external
magnetic field $\bBf$, produced by a ``free'' current density $\bJf$ which
lies entirely outside the system. We have
\begin{equation}
\bJ = \bJb + \bJf \, , \qquad 
\bB = \bBb + \bBf \, .
\end{equation}
The relation between the current and corresponding
magnetic field is given by
Amp\`ere's law,
\begin{equation}
\bnabla \times \bB = \bJ\, ,  \ \quad
\bnabla \times \bBb = \bJb\, , \ \quad \bnabla \times \bBf = \bJf \, .
\label{Bbf}
\end{equation}

As in the previous section, we separate out which terms in the Hamiltonian are
part of the material, from those which
are part of the external apparatus. For the
electrostatic case,
the energy needed to create the external field
is $(1/2)\int d \bx\, \bEf^2$,
\textit{independent of the state of the system}, so it was simple to
subtract this off to get the Hamiltonian of the material. For the
magnetostatic case the situation is more complicated since a change in the
magnetic field produced by
the material generates, according to Faraday's law,
a back EMF, $\bE$, in the coil producing the
external field, and so work has to be done to \textit{maintain} the external
current and field.
This work is done at a rate 
\begin{equation}
{{dW}\over{dt}} = -\int d\bx \, \bJf \cdot \bE \, .
\end{equation}
Using $\bnabla \times \bBf = \bJf$, integrating by parts, and using 
$\bnabla \times \bE = - \partial_t \bB,$ we obtain 
\begin{equation}
\delta W = \int d\bx  \, \bBf \cdot \delta \bB \, .
\end{equation}
Hence the total work done to maintain the external field if the total field
increases from 0 to $\bB$ (due to changes in the system) is
\begin{equation}
W = \int d\bx  \, \bBf \cdot \bB \, .
\label{workdone}
\end{equation}
This is part of the energy of the external apparatus (since it is that which
maintains the external current) and so should be \textit{subtracted}
from the total energy when determining the energy of the system. In addition,
we need to subtract
the magnetic field energy due entirely to
the external field, which is $(1/2) \int d\bx\,
\bBf^2$ from Eq.~(\ref{simpleham}) and $- \int d\bx \, \bBf^2$ from
Eq.~(\ref{workdone}).  The net effect of this is to \textit{add} $(1/2) \int d
\bx \bBf^2$.

Hence, from Eq.~(\ref{simpleham}), the Hamiltonian of the magnetic material
is~\cite{narayan,fisher}
\begin{equation}
\Hamm = \Ham_0\left(\{\bp - q \bA\}\right) +
{1\over 2} \int d\bx \, (\bB - \bBf)^2 \, .
\label{magham}
\end{equation}
where $\Ham_0\left(\{\bp - q \bA\}\right)$
is the Hamiltonian of the material in the absence of a magnetic
field except that the momentum $\bp$ of a particle is replaced by $\bp - q
\bA$ and for each spin $\bs$ there is an additional
energy $-\gamma\bs \cdot \bB$. (To avoid too cumbersome a notation we do not explicitly
indicate the dependence of $\Ham_0$ on the spins.)

Whereas the electric field in the electrostatic is not a dynamical variable,
but is completely determined by the charge density, the magnetic field
\textit{is} a dynamical variable. More precisely, to ensure $\bnabla \cdot \bB
= 0$, we write $ \bB = \bnabla \times \bA $
and treat $\bA$ as a dynamical variable 
in some gauge, e.g. $ \nabla \cdot \bA = 0 $
which is known as the Coulomb gauge. The free energy of the magnetic material
is given by
\begin{equation}
\Fm = - k_B T \ln \Tr e^{-\beta\Hamm} \, ,
\end{equation}
where the trace is over both the state of the system and the
vector potential in the Coulomb gauge.

It is conventional to describe the response of a material to a magnetic field by
its magnetization $\bM$, where $\bnabla \times \bM = \bJb \,$ and to define 
the auxiliary field $\bH$ as $\bH = \bB - \bM$ so that $\bnabla\times\bH = \bJf\, .$
Analogous to our discussion of polarization in electrostatics, by comparing 
with Eq.(\ref{Bbf}) we see that the transverse parts of $\bM$ and $\bH$ satisfy
\begin{equation}
\bM^T = \bBb \qquad\qquad \bH^T = \bBf \,.
\label{MT}
\end{equation}
However, the longitudinal part of $\bM$ and $\bH$ are not specified, unless 
$\bM$ is a local function of $\bB.$ When the magnetization is 
a non-local process, as in a
superconductor, the definition of the longitudinal part of $\bM$ is genuinely
ambiguous~\cite{Landau}.

Let the external magnetic field $\bBf$ be changed by $\delta \bBf =
\bnabla\times \delta \bAf,$
and calculate the change in free energy. Although $\bB$ 
and all parameters of the material will respond to $\delta\bBf,$
they cause no change in $\Fm$ to first order (as in the electrostatic case). 
Therefore
\begin{eqnarray}
\delta \Fm  & = &  \int d\bx \, \left[ \bnabla \times (\bBf - \bB)
\right] \cdot \delta \bAf  \nonumber \\
& = &   -\int_V d\bx \, \bJb \cdot \delta \bAf 
\nonumber \\
& = &  
-\int_V d\bx \, \, \bM \cdot \delta \bBf  \, ,
\label{dFdB}
\end{eqnarray}
where the second step used Amp\`ere's law,
Eq.~(\ref{Bbf}), and the third step an integration by parts with 
$\bnabla \times \bM = \bJb \, .$
Equation (\ref{dFdB}) is given in de Gennes\cite{degennes} and
Kittel\cite{Kittel}.
As in the electrostatic case, this 
expression can also be used to find the work done when moving a magnetic
material in an external magnetic field\cite{Kittel}. 

Next we discuss other choices that appear in the literature for the
Hamiltonian and corresponding free energy.
To a good approximation we can replace the
trace over the vector potential by the vector potential field
which minimizes the free
energy. This amounts to ignoring thermal fluctuations in the vector potential,
similar to the neglect of quantum fluctuations in the radiation field that is
customary in elementary quantum mechanics~\cite{Bethe}.

Defining 
$F_0 = - k_B T \ln \Tr^\prime e^{-\beta\Ham_0}$
we have 
\begin{equation}
\delta F_0 = \delta \Fm - \int d\bx \, \, (\bB - \bBf)\cdot(\delta\bB - \delta\bBf).
\end{equation}
Since $\delta\bB - \delta\bBf$ is transverse, from Eq.~(\ref{MT}) we can replace
$\bB - \bBf$ with $\bM.$ Using Eq.(\ref{dFdB}), we obtain
\begin{equation}
\delta F_0 
=-\int_V d\bx \, \bM \cdot \delta \bB
\,
\label{dF0dB}
\end{equation}
which can be used as the thermodynamic definition of the 
magnetization.

Some authors, e.g. Callen~\cite{callen} and Pippard~\cite{pippard},
take the Hamiltonian of the system to be
\begin{equation}
\Hammprime = \Ham_0\left(\{\bp - q \bA\}\right) + {1\over 2} \int d\bx \,
\left[\bB^2 - \bBf^2 \right] \, ,
\label{magham2}
\end{equation}
rather than Eq.~(\ref{magham}). Equation (\ref{magham2}) counts
the energy to initially set up the magnetic field,
$(1/2)\int d\bx \, \bBf^2$, to be part the external apparatus, but the 
work done by the external currents to \textit{maintain} the external field (if
some change is made to the magnetic material) is regarded as energy of the
\textit{material}.
Comparing Eq.~(\ref{magham2}) with
Eq.~(\ref{magham}), and using Eq.~(\ref{MT}), one can see that
the free energy $\Fmprime$ corresponding to $\Hammprime$ is given
by\cite{Kittel,notation}
\begin{equation}
\Fm  =  \Fmprime - \int_V d\bx \, \bM \cdot \bBf  \, .
\label{legendre_mag}
\end{equation}
Hence, from Eq.~(\ref{dFdB}), we have\cite{Kittel,pippard,callen}
\begin{equation}
d \Fmprime = \int_V d\bx \, \bBf \cdot \delta \bM \, .
\label{dFpdB}
\end{equation}

The last free energy we consider in this section, is the combined
free energy
of the external apparatus and the magnetic material. From
Eq.~(\ref{simpleham}) this is given by
\begin{equation}
\Fcomb = F_0 + {1\over 2} \int d\bx\, \bB^2 \, .
\end{equation}
From Eq.~(\ref{dF0dB}) a small change in the fields gives
\begin{equation}
\delta \Fcomb    =  \int d\bx\, \left[-\bM + \bB \right] \cdot \delta \bB =
\int d\bx\, \bH \cdot \delta \bB \,
\label{dFsadB}
\end{equation}
Equation (\ref{dFsadB}) is
the result given in the E\&M books such as Jackson~\cite{jackson}. 

\begin{table*}
\begin{center}
\begin{tabular}{|l|l|}
\hline\hline
\hspace{3.4cm}Hamiltonian & \hspace{0.8cm}Free energy change \\
\hline\hline
$ \displaystyle \Hamm = \Ham_0(\{\bp - q\bA\}) +
\smfrac{1}{2}\int d\bx\, \bBb^2  $ &
$\displaystyle \delta \Fm = -\int_V d\bx\, \bM \cdot \delta
\bBf$~[\onlinecite{Kittel,degennes}] \\
\hline
$ \displaystyle \Hamm' = \Ham_0(\{\bp - q\bA\})  +
\smfrac{1}{2} \int d\bx\, (\bBb^2 + 2\bBb \cdot \bBf) $  &
$\displaystyle \delta \Fm' = 
\int_V d\bx\, \bBf \cdot \delta \bM $~[\onlinecite{Kittel,pippard,callen}] \\
\hline
$\displaystyle \Hcomb = \Ham_0(\{\bp - q\bA\})  +
\smfrac{1}{2}\int d\bx\, (\bBb^2 + 2\bBb \cdot \bBf + \bBf^2)$
& $\displaystyle \delta \Fcomb  = 
\int d\bx\, \bH \cdot \delta \bB $~[\onlinecite{jackson}] \\
\hline
$\displaystyle \Ham_0(\{\bp - q\bA\})  $ &
$\displaystyle  \delta F_0 = - \int_V d\bx\, \bM \cdot \delta \bB$ \\
\hline\hline
\end{tabular}
\caption{Different free energies for the magnetostatic case.
We have written the magnetic field piece of the energy in
terms of the separate contributions from the bound and free currents.
In this way we clearly see that
the different Hamiltonians and free energies just
differ as to which pieces of the magnetic field energy,
$(1/2)\int d\bx\, (\bBb^2 + 2 \bBb \cdot \bBf + \bBf^2)$, are included.
The first entry in the table is
the most useful since it includes all the terms in the
Hamiltonian associated with the system. Furthermore,
the expression for $\delta \Fm$ also gives the work done on the material if
it is moved in the external field.
\label{table_mag}
}
\end{center}
\end{table*}

\medskip 

Eqs.~(\ref{dFdB}), (\ref{dF0dB}), (\ref{dFpdB})
and (\ref{dFsadB}) describe
different thermodynamic
derivatives in the presence of an magnetic field, each of which
involves a different
free energy
depending upon which pieces of the magnetic field energy
$(1/2)\int d\bx\, (\bBb^2 + 2 \bBb \cdot \bBf + \bBf^2)$ are included.
These are summarized in Table~\ref{table_mag}.
In situations where the definition of $\bM$ (and hence $\bH$)
is ambiguous one can replace $\bM,$ $\delta\bM$ and $\bH$ with
$\bBb,$  $\delta\bBb$ and $\bBf$ respectively in the last column of 
Table~\ref{table_mag}, after extending the integrals to cover all
space. This relies on Eq.~(\ref{MT}) and the fact
that $\delta\bBf,$  $\bBf$ and $\delta\bB$ are transverse. 
Of the free energies in Table~\ref{table_mag},
$\Fm$ is the most useful since it includes everything
associated with the magnetic material: (i) the
part of the Hamiltonian not involving 
the magnetic field (except that we replace $\bp \to \bp - q \bA$ and add
$-\gamma\bs \cdot \bB$ for each spin) 
and (ii) the long-range dipole-dipole
interaction between
magnetic moments in the material. 

\section{Conclusions}
The main results of the paper are summarized in Tables~\ref{table_el} and
\ref{table_mag}.
For statistical mechanics applications,
we have argued that the most appropriate Hamiltonian for the \textit{system}
(neglecting the Hamiltonian of the external apparatus) is the first line of
these tables, i.e.
\begin{equation}
\Hamd =  \Ham_0 + 
{1 \over 2} \int d\bx \, \left[ \bE^2 - \bEf^2 \right] 
\end{equation}
for the electric case and
\begin{equation}
\Hamm = \Ham_0\left(\{\bp - q \bA\}\right) + {1\over 2} \int d\bx \, (\bB -
\bBf)^2 \, .
\end{equation}
for the magnetic case. The corresponding free energy changes are
\begin{equation}
\delta \Fd = -\int_V d\bx\,  \bP \cdot \delta \bEf 
\label{final_el}
\end{equation}
and
\begin{equation}
\delta \Fm  = 
-\int_V d\bx \, \, \bM \cdot \delta \bBf  \, . 
\label{final_mag}
\end{equation}
The relation between the first line of the tables and the second is given, in
each case, by a standard Legendre transform, see Eqs.~(\ref{legendre_el}) and
(\ref{legendre_mag}).
Unfortunately, Eqs.~(\ref{final_el}) and (\ref{final_mag}) do
\textit{not} figure prominently in statistical mechanics textbooks. Reference
[\onlinecite{Kittel}] is a notable exception, but unfortunately is not in
print. By
contrast, for problems in electricity and magnetism we are generally
interested in the total work done (on the \textit{free} charges and currents),
which is given by the third line of the tables, and fortunately, these 
\textit{are} the results quoted in E\&M books.

It is interesting to note that although the final results for the free energy
changes in the electric and magnetic cases are exactly analogous to each
other, there are differences in the Hamiltonians from which these results are
derived. The first difference is that the field energy involves 
$(1/2)\int d\bx\, (\bE^2 - \bEf^2) = \int d\bx\,(\bEb^2/2 - \bEb\cdot\bEf)$ 
in the electric case as opposed to
$(1/2)\int d\bx\, (\bB - \bBf)^2 = (1/2)\int d\bx\, \bBb^2.$ 
The extra piece in the magnetic case, $\int d\bx\, (\bBf^2 - \bBf \cdot \bB),$
comes from the work done by the external
sources in the magnetic case to maintain the external currents against the back
EMF produced by a change in the field of the system. The second difference is
that whereas $\Ham_0$ does not have an explicit dependence on the electric
field (though it depends on it implicitly, since $\bE$ is tied to the charge
density through Gauss' law, Eq.~(\ref{divE})), it does depend
explicitly on the magnetic field, since one has to replace the
velocity $\dot{\bx}$ by $\bp - q \bA$ where $\bp$ is the canonical momentum.
For the case of 
magnetism due to spins, $\bs_i$, there is also an explicit interaction
$-\gamma \sum_i \bs_i \cdot \bB$ in $\Ham_0$.
Classically, one can change variables from $\bp$ to 
$\bp' = \bp - q \bA,$ there is no spin magnetism, and the field energy only 
involves $\bBb,$ so there is
no magnetism~\cite{Bohr}. However, since in quantum mechanics there are 
constraints on $\bp$ and not $\bp',$ for example that the angular momentum
${\bf r}\times\bp$ must be a multiple of $\hbar,$ there is a 
coupling between the dipole moment of the material and the external field 
from the $\Ham_0(\{\bp-q\bA\})$ part --- or the spin term ---
of the Hamiltonian.

It is useful to compare the \textit{equilibrium} free energies
we have discussed so far with the Landau free energy function
which forms the basis of
Landau's theory of second order phase transitions\cite{landau_sm}, and which
is applicable \textit{away from equilibrium}. 
The
partition function, required in evaluating the free energy through
statistical mechanics, is written (for the magnetic case) as
\begin{eqnarray}
Z(\bBf) & = &  \sum_\bm \Tr'\exp[-\beta\Hamm] \nonumber \\
& = & \sum_\bm \exp[-\beta \FLand(\bm, \bBf)] \, ,
\label{landau}
\end{eqnarray}
where $\Tr'$ is a restricted trace, carried out over states with a
specified value of total magnetic moment $\bm,$ and the Landau free energy
$\FLand$ is defined through this equation. In Eq.~(\ref{landau}),
the sum is dominated by the value of $\bm$ which minimizes $\FLand$
at fixed $\bBf$ (assumed independent
of $\bx$ here); this is the equilibrium magnetization at $\bBf.$
Denoting this equilibrium value of $\bm$ by $\bm^*,$ we have~\cite{ma}
\begin{equation}
\FLand(\bm^*,\bBf) = \Fm(\bBf) \, .
\end{equation}

For the case of spin magnetism, Eq.~(\ref{landau}) simplifies further. As
seen in Eq.~(\ref{simpleham}),
the coupling between the external field and the system
occurs
only through the term $-\bm\cdot\bBf,\ (\bm = \gamma \sum_i \bs_i)$,
the part of $-\bm\cdot\bB$ involving the external field. Therefore
\begin{equation}
\FLand(\bm, \bBf) = F(\bm) -\bm\cdot\bBf
\end{equation}
where $F(\bm)$ does not depend on the externally applied magnetic
field. In this case, comparing
with Eq.~(\ref{legendre_mag})
we see that\cite{ma}
\begin{equation}
F(\bm)=\Fm' \, .
\end{equation}
A similar
process can also be carried out for the electrostatic case, since the
coupling term in $\Hamd$ is $-\int d\bx\, \bEb\cdot\bEf = - \int d\bx\,
\bP\cdot\bEf.$ However, for orbital magnetism, $\bBf$ cannot be separated
out cleanly like this due to the diamagnetic term, and one is limited to 
Eq.~(\ref{landau}) if the diamagnetic contribution is significant.

Finally, we observe that the `external' parts of the Hamiltonian, that were
discarded in going from Eq.~(\ref{simpleham}) to Eqs.~(\ref{HamdE}) and 
(\ref{magham}), were $(1/2)\int d\bx\, \bEf^2$ and 
$-(1/2)\int d\bx\, \bBf^2.$ The sign reversal between the two leads to the
well known result~\cite{Feynman} that free charges move to 
minimize the energy of the electric field (the position of the charges in a 
conductor can be viewed as such a minimization problem, with constraints)
whereas loops carrying free current (held constant by batteries) move
to maximize the energy stored in the magnetic field. 

We hope that a brief discussion of the points discussed here will eventually
find their way into textbooks on statistical mechanics so that future students
will be less confused about the free energy of a system in an electric or
magnetic field than were the authors of this paper.

We would like to thank D.~Belanger for a critical reading of the manuscript.


\begin{thebibliography}{99}
\bibitem{notation}
There does not seem to be any standard notation for the free energies that we
call $\Fm$ and $\Fm'$. For example Kittel\cite{Kittel} calls them
$F_A$ and $F_B$.
Some authors use the terms ``Gibbs'' and
``Helmholtz'' free energies, but we prefer to reserve these terms for free
energies which differ by a factor of $P V$.
\end{thebibliography}
\end{document}